
\def\cm{{\rm\thinspace cm}}
\def\erg{{\rm\thinspace erg}}

\def\hmpc{\thinspace h^{-1}\ {\rm Mpc}}

\def\s{{\rm\thinspace s}}


\def\ergpcmsqps{\hbox{$\erg\cm^{-2}\s^{-1}\,$}}

\def\ref{\par \noindent\hangindent=3pc \hangafter=1}

\def\etal{{\it et al. \thinspace}}

\def\spose#1{\hbox to 0pt{#1\hss}}
\def\approxlt{\mathrel{\spose{\lower 3pt\hbox{$\sim$}}\raise 2.0pt\hbox{$<$}}}
\def\approxgt{\mathrel{\spose{\lower 3pt\hbox{$\sim$}}\raise 2.0pt\hbox{$>$}}}
\mathchardef\twiddle="2218

\def\multleft#1{\hbox to size{\vbox {\halign{\lft{##}\cr #1}}\hfill}\par}
\def\multright#1{\hbox to size{\vbox {\halign{\rt{##}\cr #1}}\hfill}\par}
\def\HEAO1{{\sl HEAO--1}}
\def\Ginga{{\sl Ginga}}
\def\Mdot{\rlap{\raise 9pt\hbox{\hskip 5pt\char'56}}{\hbox{$M$}}}
\def\Mdot{\hbox{$\dot M$}}

\def\today{\ifcase\month\or January\or February\or March\or April\or May\or
June\or July\or August\or September\or October\or November\or December\fi
\space\number\day, \number\year}
\def\<{\thinspace}

\font\big=cmr10 scaled\magstep2
\font\bigbf=cmbx10 scaled\magstep2

\font\bigsl=cmsl10 scaled\magstep2

\tolerance=10000

\magnification=\magstep1
\baselineskip=0.7truecm
\hsize=6.2truein
\vsize=9.0truein
\hoffset=0truein
\voffset=0truein
\parindent=0.5truein
%
%

\noindent {\bigbf Nearby Galaxies and the {\bigsl Ginga}\ X--ray Background}

\vskip 2truecm

\noindent {\big F.J. Carrera$^1$, X. Barcons$^{2}$, J.A. Butcher$^3$,
A.C. Fabian$^4$, O. Lahav$^4$, G.C. Stewart$^3$, \& R.S. Warwick$^3$}

\vskip 1.5truecm

\noindent $^1$ {\it Mullard Space Science Laboratory -- University College
London, Holmbury St. Mary, Dorking, Surrey, RH5 6NT}

\smallskip

\noindent $^2$ {\it Instituto Mixto (Consejo Superior de Investigaciones
Cient\'\i ficas--Universidad de Cantabria), 39005 Santander, Spain}

\smallskip

\noindent $^3$ {\it X--ray Astronomy group, Physics Department, University of
Leicester, Leicester LE1 7RH}

\smallskip

\noindent $^4$ {\it Institute of Astronomy, Madingley Road, Cambridge CB3 0HA}

\vskip 2truecm

\noindent Submitted to {\it Monthly Notices of the Royal Astronomical Society}

\vskip 2truecm

\leftskip 2.5truecm

\noindent {\bf ABSTRACT}

\noindent We present here the results of cross--correlating the X--ray
background measured by {\sl Ginga} in the 2--10~keV band
with several catalogues of extragalactic objects.
Positive signals with an amplitude of a few per cent have been found for some
catalogues implying that some fraction of the X-ray background is produced
either
by the class of catalogued sources or by other classes spatially related to
them. Detailed X--ray background simulations have
been used to assess the significance of the results and, for the first time,
the full angular shape of the cross--correlation. The inferred X--ray volume
emissivity in the local Universe, $j_0$, has been estimated for two galaxy
catalogues
(UGC and IRAS) for which the cross--correlation is highly significant. We
obtain  $j_0= (0.74\pm 0.07)$ for UGC and $j_0= (1.15\pm 0.10)$ for IRAS,
 in units of
$10^{39}\thinspace h\ {\rm erg\ s^{-1} \ Mpc^{-3}}$.
Extrapolating this result back to $z\sim1-4$ leads to the conclusion that
$\approxlt 10-30$ per cent of the X--ray background could be produced by a
non--evolving population of galaxies. These values are shown to be consistent
with upper limits on the Auto Correlation Function derived here.

\leftskip 0cm

\vfill\eject

\beginsection 1 INTRODUCTION

\noindent The puzzle of the origin of the X--ray background (XRB) still remains
unsolved, although it is becoming apparent that no single class of
`miraculous' sources is able to satisfy simultaneously all the observational
constrains accumulated over the last 30 years (see Fabian and Barcons 1992
for a review).

The fraction of the XRB directly resolved into sources  in the `soft' band
($<2$~keV) is about 50
per cent in the deepest ROSAT surveys (Hasinger \etal 1993,
Branduardi--Raymont \etal 1994). The brightest of these sources (i.e., those
with 0.5--2~keV fluxes $\approxgt 10^{-14}\, {\rm erg}\, {\rm cm}^{-2}\, {\rm
s}^{-1}$) are
expected to be quasars with $z\leq 2.5$ (based on the optical identification
work of Shanks \etal 1991). Optical
identification programs are being carried out in order to establish the nature
of the sources at and below that flux.  On the other hand, fluctuation studies
of these Deep {\it Rosat} fields (Hasinger et al. 1993, Barcons et al. 1994)
reveal the imprint of $\sim 1000$~sources deg$^{-2}$, `resolving' a fraction of
$\sim 70$ per cent of the soft XRB. The nature of the sources at fluxes
$\sim 10^{-15}\, {\rm erg}\, {\rm cm}^{-2}\, {\rm s}^{-1}$ will not be known
until X--ray telescopes more sensitive than {\it Rosat} become available.

The lack of imaging devices in the `hard' band (2--10~keV) has made the
directly resolved fraction of the XRB in that band much smaller, only a few per
cent (Piccinotti \etal 1982). Integrating and extrapolating the obtained
luminosity
functions, clusters of galaxies would contribute $\sim4$ per cent of it and AGN
(mostly Seyfert galaxies) $\sim20$ per cent. Although soon ASCA, and
SPECTRUM--X--$\Gamma$ in the future, are expected to give some insight into
this problem, useful information is available contained in the angular
distribution of the intensity of the XRB (fluctuations and anisotropies).

Fluctuation studies using {\sl Ginga} data (Butcher \etal 1994) have shown that
the $\log\ N-\log\ S$ curve is consistent with an euclidean shape (sources per
unit flux and unit solid angle $n(S)\propto S^{-2.5}$)  down to $\sim 10^{-12}
\ergpcmsqps$ (2--10~keV). Recent XRB anisotropies and excess fluctuations
studies have shown that only 60--70~per cent of the `hard' band XRB can be
produced by sources clustering on scales $\sim 6 \hmpc$ ($H_0=100\thinspace h\,
{\rm km\,s^{-1}\,Mpc^{-1}}$) (see Danese \etal 1992 for a recent review). This
implies that the obvious solution of just `adding' more QSOs until the XRB
saturates does not trivially work.

A complementary way of facing the problem of the origin of the XRB is to search
for cross--correlations of the XRB intensities with positions of known sources
in existing catalogues.

We define the cross--correlation function (CCF) as

$$W_{\rm xg}(\theta)={\langle\delta N\cdot\delta I\rangle_\theta \over \langle
N\rangle\langle I\rangle},\eqno(1)$$

\noindent $\delta N$ being the fluctuation in projected galaxy density
 and $\delta I$ the XRB deflection,
 $\langle N\rangle$ and $\langle I\rangle$ being the
mean projected galaxy density and XRB intensity, respectively (see Sections 2
and 3 for
definitions). $\langle\ \rangle_{\theta}$ means an average among all those
pairs distant an angle $\theta$.

Jahoda \etal (1991) obtained a value of $3\times
10^{-3}$ for the zero--lag  cross--correlation function (CCF) of the 2--10~keV
XRB intensities (as observed by HEAO--1) with ESO and UGC galaxies. Jahoda
\etal (1992), using this value,  concluded that a sustantial fraction of the
`hard' XRB could be produced by a non--evolving
population of  X--ray sources.

Lahav \etal (1993) performed a more sophisticated but preliminary analysis of
{\sl Ginga} and HEAO--1 data at zero--lag. They
arrived at the conclusion that  $\sim$30--50 per cent of the `hard' XRB could
be
produced by local non--evolving X--ray sources. Their analysis (unlike that by
Jahoda \etal 1992) took into account the clustering of the sources, which, as
shown there and in Section 4, can significantly reduce the values of the
emissivity deduced from the observed CCFs.

Miyaji \etal\ (1994) obtained  $j_0=(0.9\pm0.2)
\times 10^{39}\thinspace h\ {\rm erg\ s^{-1} \ Mpc^{-3}}$
using the zero--lag CCF of the whole sky \HEAO1
XRB observations with two catalogues of IRAS galaxies.
They also used a sample of AGN detected by \HEAO1 (Grossan 1992) to study the
spatial correlation
function and the selection function of sources detected both in X--rays and
infrared wavelengths.

In the present paper we have analyzed the CCF obtained from a set of {\sl
Ginga} scan data (see Section 2.1) with nine catalogues of extragalactic
sources (listed in Section 2.2). We have used, for the first time, the full
angular shape of the CCF up to 5 degrees to gather information about possible
sources of the 2--10~keV XRB and their relation to the sources in the
catalogues. We have also taken into account the limits imposed by the observed
auto--correlation function (ACF) of the XRB over the same region of the sky.

In Section 3 we show how the CCF of the XRB with the galaxy catalogues is
obtained. We also discuss the significance of the signals seen, basing our
arguments on detailed simulations of the X--ray sky and the galaxy catalogues
in the region under study
(Section 3.2).

Section 4 is devoted to the theoretical framework in which the CCF is modelled.
We give the expression of the expected CCF as a function of the properties of
the sources (selection functions, spatial correlation functions and
emissivities) and the collimator profile (`beam').

Our results are used in Section 5 to estimate the volume emissivity of X--ray
sources related to a local population of galaxies. The implications of
these results are then addressed in Section 6.

Finally, in Section 7 we summarize our results.

\beginsection 2 THE DATA

\smallskip

\noindent {\bf 2.1 The X--ray data}

\medskip

\noindent Our X--ray sample consists of 5 strips over the North Galactic Pole
observed in
scan mode with the Large Area Counter (LAC) on board of {\sl Ginga}. Every
strip was scanned several times by the satellite, accumulating counts in 16 s
time intervals, which correspond to $\sim 0.2$~degrees angular separation. As
consecutive scans in the same strip do not exactly overlap, we actually have
measurements of the brightness of the X--ray sky with relative separations
smaller than that distance. However, as the collimator profile  is only well
determined in 0.1~deg steps, we have binned the observed counts in 0.2~deg wide
bins. This binning also helps in improving the signal--to--noise ratio.

A more thorough description of the data and their reduction process is given
in Carrera \etal (1993), our data being a subset of those used in that work.
We only stress here that the energy band used ({\sl Ginga} LAC Pulse Heigth
Analyzer -PHA-  channels 8--20, corresponding roughly to 4--12~keV), and the
high galactic latitude of the sample ($\mid b\mid> 30\,{\rm deg}$),
lead to a small contamination by diffuse gas from our galaxy. Furthermore, the
extension of this
region is small and the angle between it and the dipole
direction is 86~degrees, making insignificant any dipole
correction because of the Compton--Getting effect ({\it e.g.}, Boldt 1987).
Given that our sample is at a supergalactic latitude $\sim 40$~degrees, the
supergalactic plane do not pose any problems of fair sampling on it.

As the region of the sky scanned is also away from known hard X--ray sources
(those of Piccinotti \etal 1982), we have a sample of extragalactic diffuse XRB
deflections or fluctuations $\delta I$
$\equiv I-\langle I\rangle$, $\langle I\rangle$ being
the mean XRB intensity.

\medskip

\noindent {\bf 2.2 The catalogue samples}

\medskip

\noindent A total of 7 galaxy samples have been used in this work, together
with the
Abell catalogue of clusters of galaxies. The sources from each catalogue with
Right Ascension
$(\alpha)\ \in [210,270]$ degrees and
Declination $(\delta)\ \in [30,70]$ degrees have been selected and
their positions used to define a set of fluctuations in projected galaxy
density
($\delta N$) in the following way: at the same positions and with the same
collimator orientation as in the real XRB observations, a series of galaxy
number density ($N$) `observations' have been performed. Then they are binned
in 0.2~deg wide bins (again  in the same way as in the X--ray observations)
and their mean value ($\langle N\rangle$) subtracted to give the
fluctuations $\delta N$.

The number of sources used from each catalogue and the value of
$\langle N\rangle$ for each one of them (in ${\rm sources\ beam^{-1}}$) are
given in Table 1.

Briefly the catalogues are as follows:

\item{-} IRAS: the Meurs and Harmon (1989) catalogue that comprises
infrared galaxies selected (using purely colour criteria)
from the IRAS Point Source Catalogue (PSC) with
$60\,\mu {\rm m}$ fluxes $>0.7\,{\rm Jy}$. This catalogue was also used
by Miyaji \etal\ (1994).

\item{-} UGC: the Uppsala General Catalogue (Nilson 1973) of optical galaxies,
selecting those whose major angular diameters are greater than 1~arcmin.

\item{-} IRASQDOT: IRAS QDOT catalogue taking one in six sources out of a
galaxy subsample of the PSC (Rowan--Robinson \etal 1990). 97 per cent of the
2163 sources in this catalogue ($60\,\mu {\rm m}$ fluxes $>0.6\,{\rm Jy}$ and
$\mid b\mid>10^{\circ}$) have been spectroscopically identified.

\item{-} CfA: Centre for Astrophysics redshift catalogue of galaxies
(Huchra \etal 1983), which
happens to partly overlap with the region of the sky observed in X--rays.

\item{-} Abell: positions of Abell (1958) clusters. The richnesses of the
selected clusters are between 0 and 4, most of them having
Richness class 0 or 1.


\item{-} IRASZ: Strauss \etal (1990) sample of IRAS PSC galaxies complete
down to $60\,\mu {\rm m}$ fluxes $>1.936\,{\rm Jy}$, all with
$\mid b\mid>5^{\circ}$. Redshifts are available for these sources. There are
two known AGNs in this catalogue within our sample region; including or
excluding them does not make any appreciable difference to the results
discussed below (neither does it to those of IRASZS or IRASZSVOL below).

\item{-} IRASZS: a selection among galaxies in the IRASZ sample, choosing
those with $500<cz_{\rm Local Group}<8000\ {\rm km\ s^{-1}}$.
This is done (Yahil \etal 1991) to avoid both local galaxies with possible
peculiar velocities and very distant ones, for which the selection function
(probability of a source being detected; see Section 4) is very small. This
catalogue was also used
by Miyaji \etal\ (1994).

\item{-} IRASZSVOL: this comprises the same sources as in IRASZS, but
we weighted them by the inverse of the selection function at their distances.
The idea behind this weighting is to 'compensate' for the sources missing in
the sample (the farther away, the more likely to be missing, the smaller the
selection function and the larger the weight).

\beginsection 3 THE OBSERVED CROSS--CORRELATION FUNCTIONS

\smallskip

\noindent {\bf 3.1 Measurement of the Cross Correlation functions}

\medskip

\noindent In Section 1 we defined the CCF as

$$W_{\rm xg}(\theta)={\langle\delta N\cdot\delta I\rangle_\theta \over \langle
N\rangle\langle I\rangle}$$

To evaluate $\langle I\rangle$ we have used the spectrum of the XRB obtained by
Marshall \etal (1980). It leads to $\langle I \rangle=5.7\times 10^{-8}\,{\rm
erg\,cm^{-2}\,s^{-1}\,sr^{-1}}$ in the 2--10~keV band, which corresponds to
$\sim 8\,{\rm ct\ s^{-1}\ beam^{-1}}$ in the LAC PHA channels 8 to 20.

The values of $W_{\rm xg}(0)$ in 1~deg bins derived from the different samples
are noted in Table 1. The UGC--XRB CCF value is larger than the one obtained by
Jahoda \etal (1991) ($W_{\rm xg}(0)\sim (2-5)\times 10^{-3}$). Our CCF value
for  the IRAS and IRASZS are also larger than their equivalents in Miyaji
\etal\ (1994).  This is hardly surprising, since the CCF is heavily dependent
on instrument characteristics ($I$ is measured in counts beam$^{-1}$ and
$N$ in galaxies beam$^{-1}$). For example, the {\sl Ginga} LAC probes deeper
in source counts than \HEAO1 A2 due to its smaller beam size
($2^\circ\times1^\circ$ and $3^\circ\times1.5^\circ$ respectively). The
different beam sizes also have a geometrical effect that contributes to make
the signal different (see Section 4).

We have plotted $W_{\rm xg}(\theta)$ in 1~deg wide bins for the UGC and IRAS
samples in Figs. 1$a$ and 1$b$, respectively (filled dots). The error bars are
from the standard deviation in each bin. As we will
show in Section 4, CCF signals (such as those seen in Figs. 1$a$
and 1$b$)
are expected due to the finite size of the {\sl Ginga} LAC collimator
($\sim 1^\circ\times 2^\circ$) and/or to the spatial clustering  of
the X--ray sources with the galaxies in the samples (see also Lahav \etal
1993).

\medskip

\noindent {\bf 3.2 Significance of the Cross Correlation Signal}

\medskip

\noindent We have simulated both catalogue samples (simply by reshuffling at
random the
positions of the objects in them) and X--ray fluctuations to assess the
significance of our results. By obtaining the CCFs of these random
samples, we can test if purely geometrical effects or random positioning
could give rise to the observed CCFs.

The simulated X--ray samples have been obtained by uniformly distributing
X--ray sources following the $\log\ N-\log\ S$ curve obtained by
Butcher \etal (1994) (see also Carrera \etal 1993 for a complete description of
the simulation process).

All these simulated populations (both X--ray and catalogue) have then been
folded through the {\sl Ginga} LAC collimator and the corresponding
deflections have been obtained as described in Section 2.2.

The values in the column labelled $W_{\rm xg,sim}(0)$ of Table 1 are obtained
using 100 simulations of every galaxy sample and 100 simulated XRB samples: the
quantities quoted are 2$\sigma$ upper limits (they are the higher of the two
values that encompass 95 per cent of the simulated $W_{\rm xg}(0)$ for each
sample). The CCFs from the samples UGC, IRAS and IRASZS are well above those
upper limits,  showing that they are $>>2\sigma$ significant, and that they do
not come from some chance positioning of the sources in the samples or the
X--ray data, nor from some geometrical effect due to the collimator or the way
the CCFs have been obtained. The signals for the CfA and IRASZ samples are just
about at the $2\sigma$ significance level.

These differences between the CCFs from different samples reflect their
different densities, flux limits and clustering properties, and, of course,
their different percentages~of/clustering~with X--ray sources giving rise to
the XRB.

In Figs. 1$a$ (UGC) and 1$b$ (IRAS) the medians of the simulated CCFs ($
W_{\rm xg,sim}(\theta)$) are shown as solid lines, as well as values
encompassing
68 per cent (dashed lines) and 95 per cent (dotted lines) of the simulations.

To test whether most of the signal is due to a few sources just unresolved in
X--rays, we also calculated the CCF (both real and simulated) excluding the
five per cent higher X--ray deflections. Both UGC and IRAS CCFs were still
above 97~per cent of the simulations, but IRASZS (and CfA and IRASZ) was well
below that limit. This, as said earlier, indicates that the CCF signal found
for IRASZS was due to a few `bright' X--ray sources just unresolved, while
the UGC and IRAS signals come from the sources in those samples as a
population. Therefore, we pay special attention to the understanding of these
signals.

The simulated CCFs have also been used to estimate the error bars for the
observed CCFs at each angular separation (from the 68 per cent limits for each
simulation). These are shown in Figs. 2$a$ and 2$b$ (see Section 5).

Having checked and established the significance of the observed CCFs,
their relationship to the properties of the (catalogue and X--ray) sources will
be outlined in next section.

\beginsection 4 THE THEORETICAL CROSS--CORRELATION FUNCTION

\noindent In order to model the CCF on the basis of the relationship between
the X--ray sources responsible for the observed fluctuations in the X--ray sky
and the catalogued optical/IR sources, let us consider two (distinct) source
populations. The first one would consist of galaxies with selection function
$P(r)$ (probability of a source being detected at a distance $r$ from the
observer) and mean spatial density $\langle n_{\rm g}\rangle$. Our second
population
would consist of  X--ray sources with local volume emissivity $j_0$ (emitted
X--ray power per unit volume) and clustered with the above galaxies, with a
spatial cross--correlation function $\xi_{\rm xg}({\bf r}_{\rm g}- {\bf r}_{\rm
x})$
(probability in excess of a purely poissonian distribution that  a galaxy and
an X--ray source lie at a distance $\mid {\bf r}_{\rm g}- {\bf r}_{\rm x}
\mid$). We will
assume, as usual,  a form $\xi_{\rm xg}(r)=(r/r_0)^{-\gamma}$ for $\xi_{\rm
xg}(r)$. $\xi_{\rm xg}(r)=0$ (or equivalently, $r_0=0$) would mean that these
two
populations are independently distributed in the sky.


Some of the X--ray sources may be actually present in the catalogue. In this
case X--ray and catalogue observations  in different directions (given by the
unit vectors ${\bf n}_{\rm x}$ and ${\bf n}_{\rm g}$ respectively) would be
correlated when
viewd through a collimator with profile $G({\bf n}_{\rm x}-{\bf n}_{\rm g})$.
If we define

$$\eta({\bf n}_{\rm x}-{\bf n}_{\rm g})=\langle I\rangle\cdot\langle
N\rangle\cdot
W_{\rm xg}({\bf n}_{\rm x}-{\bf n}_{\rm g})\eqno(2)$$
then the correlation term without taking into account the possible clustering
of the catalogued sources (poissonian term) would be

$$\eta_{\rm pois}({\bf n}_{\rm x}-{\bf n}_{\rm g})=\int dr\ r^2{j_{\rm xg}\over
4\pi r^2}
P_{\rm x}(r) \ \int d^2\Omega_n\ G({\bf n}-{\bf n}_{\rm x})G({\bf n}-{\bf
n}_{\rm g})\eqno(3)$$
where $j_{\rm xg}\ (\leq j_0)$ is the local X--ray volume emissivity of the
galaxies that are also X--ray emitters
(we have assumed that $j_{\rm xg}$ is constant within the galaxy sample
depth)
and $P_{\rm x}(r)$ their selection function. Defining
the effective catalogue depth as

$$R_*=\int dr\ P_{\rm x}(r)\eqno(4)$$
and the collimator autocorrelation function (ACF) as

$$\eta_{\rm col}({\bf n}_{\rm x}-{\bf n}_{\rm g})=\int d^2\Omega_n\ G({\bf
n}-{\bf n}_{\rm x})
G({\bf n}-{\bf n}_{\rm g})\eqno(5)$$
then

$$\eta_{\rm pois}({\bf n}_{\rm x}-{\bf n}_{\rm g})={1\over 4\pi} j_{\rm xg} R_*
\eta_{\rm col}({\bf n}_{\rm x}
-{\bf n}_{\rm g}) \eqno(6)$$

If we now allow for clustering among the two populations, the CCF is

$$\eta_{\rm clus}({\bf n}_{\rm x}-{\bf n}_{\rm g})=\int dV_{\rm x'}\ {j_0\over
4\pi r_{\rm x'}^2}
\int dV_{\rm g'}\ P(r_{\rm g'})\langle n_{\rm g'}\rangle
\ G({\bf n}_{\rm x'}-{\bf n}_{\rm x})G({\bf n}_{\rm g'}-{\bf n}_{\rm g})
\xi({\bf r}_{\rm x'}-{\bf r}_{\rm g'}) \eqno(7)$$
which, under the small angle approximation and using Eq. 5, leads to

$$\eta_{\rm clus}({\bf n}_{\rm x}-{\bf n}_{\rm g})={1\over 4\pi}j_0\langle
n_{\rm g}\rangle H_\gamma
r_0^\gamma \int dr\ r^{3-\gamma}P(r)
\int d^2\Omega_n\ \eta_{\rm col}({\bf n}-{\bf n}_{\rm x})
\mid {\bf n}-{\bf n}_{\rm g}\mid^{1-\gamma}\eqno(8)$$
where again we have assumed a constant $j_0$ over the volume of the galaxy
sample
and $H_\gamma=\Gamma(1/2)\Gamma((\gamma-1)/2)/\Gamma(\gamma/2)$, $\Gamma$ being
the gamma function.

The total CCF would then be the sum of these two terms (Eqs. 6 and 8).
If clustering is present, but its contribution neglected, the volume
emissivity is grossly overestimated (by as much as a factor of 6 in the
relevant case). Eq. 8
generalizes Eqs. 2 and 3 in  Lahav \etal (1993) paper for the non--zero lag
non--square collimator case. Note that the emissivity ($j_0$ in this paper)
was denoted as $\rho_{\rm x}$ in that paper.

For the UGC and IRAS samples
the selections functions are well determined (Hudson and Lynden--Bell 1991 and
Yahil \etal 1991 respectively).

Miyaji \etal\ (1994) constructed the spatial correlation function of a sample
of \HEAO1 X--ray sources with the IRAS catalogue, obtaining $\gamma=1.8$ and
$r_0=4 \hmpc$. They also derived the selection function for the X--ray sources
that were also in the IRAS 0.7~Jy sample (of which our IRAS sample is a
subset) $P_{\rm x}(r)$, obtaining a value of $R_*=65 \hmpc$. These are the
values of
$\gamma$,  $r_0$ and $R_*$ that we have used for our IRAS sample.

For the UGC sample we
have assumed $P_{\rm x}(r)=P(r)$ (the selection function of the X--ray emitting
galaxies in the UGC catalogue is the same as the global one) for simplicity,
since its real value is not known.
Furthermore, we have used the spatial correlation function of the optical
galaxies as the cross--correlation function of them with the X--ray sources,
{\it i.e.}, $\gamma=1.8$ and $r_0=5 \hmpc$.

Finally, in order to minimize the number of free parameters, we have assumed
$j_{\rm xg}=j_0$ (the X--ray sources are all galaxies, and they are clustered
like the galaxies of the catalogue).
The effects of this assumption are discussed in Section 6.

Taking into account all the above assumptions, from Eqs. 6 and 8 we
obtain
$$\eqalign{\eta({\bf n}_{\rm x}-{\bf n}_{\rm g})&={j_0\over 4\pi}
\biggl( R_* \eta_{\rm col}({\bf n}_{\rm x}-{\bf n}_{\rm g})\cr
&+\langle n_{\rm g}\rangle H_\gamma
r_0^\gamma \int d^3r\ r^{3-\gamma}P(r)
\int d^2\Omega_n\ \eta_{\rm col}({\bf n}-{\bf n}_{\rm x})
\mid {\bf n}-{\bf n}_{\rm g}\mid^{1-\gamma}\biggr) \cr}\eqno(9)$$

The value of $\langle n_{\rm g}\rangle$ is easily calculated taking into
account

$$\langle N\rangle=\int d^3r\ P(r)G({\bf n})\langle n_{\rm
g}\rangle=\Omega_{eff}\cdot
\int dr\ r^2P(r)\cdot\langle n_{\rm g}\rangle\eqno(10)$$
where $\Omega_{eff}=\int d^2\Omega_n\ G({\bf n})$ is the effective solid angle
 of the
collimator (in our case $5.7\times10^{-4}\ {\rm sr}$), and the values of
$\langle N\rangle$
are listed in Table 1.

The collimator profile of the {\sl Ginga} LAC is well known down to scales
$\sim 0.1$\ degrees,
but unfortunately the geometry of the X--ray  observations is quite
complex, involving many different relative positions and orientations of
the collimator. We have then obtained $\eta_{\rm col}(\theta)$ from the set of
simulations of X--ray sources previously used to estimate the significance of
the observed CCFs, and using the relation $\langle \delta I\cdot \delta
I\rangle_\theta=\langle S^2\rangle\cdot\eta_{\rm col}(\theta)$ where $\langle
S^2\rangle$ is the mean of the square of the X--ray flux $S$ as calculated from
 the
$\log\ N-\log\ S$ curve used in the simulations (see Section 3.2). In this way
we obtain a collimator ACF `tailored' to our needs, including all the
geometrical effects of our X--ray data.

Finally, the `zero' level of the CCF ({\it i.e.},
the cross--correlation expected in the absence of any true cosmic
signal) is not zero because the finite size of the sample introduces a
negative baseline in the correlation functions (Kondo 1990). This zero level
($=-W_{\rm xg}(0)/<$Number of pairs at zero--lag$>$)
has been calculated and subtracted from the observed CCF before fitting.

\beginsection 5 RESULTS: THE VOLUME EMISSIVITY

\noindent Although our angular bin size (0.2 deg) provides us with many bins,
they are hardly independent (due to the collimator extension and to the fact
that we are dealing with the product of pairs of observations each of which
appear in many pairs). Therefore, we have only kept one free parameter ($j_0$).
Its value has been obtained by fitting (using minimum least squares) the model
given in Eq. 9 to the observed CCF binned in 0.2~degree bins. This has been
done for two different angular ranges ($0.2\rightarrow 2$ deg and
$0.2\rightarrow 5$ deg) and for several values of $R_{\rm min}$,
which corresponds to the distance beyond which the selection function drops
below one ($P(r<R_{\rm min})=1$), {\it i.e.}, galaxies begin to be lost or
undetected. The results are shown in Table 2.

There is a significant difference between fits to the different angular ranges:
this is because $\eta_{\rm clus}(\theta)$ is significantly different from zero
between 2 and 5~degrees and then the fitting over the larger range is bound to
require a lower normalization ($\rightarrow$ lower $j_0$). This is shown in
Figs. 2$a$ (UGC) and 2$b$ (IRAS) where the models marked with an asterisk in
Table 2 are plotted along with the observed CCFs: the solid lines correspond to
the total CCFs, while dashed lines are the clustering terms and dotted lines
the poissonian ones. The error bars on the observed CCFs are from the
simulations of Section 3.2.

The peaks visible in Figs. 2$a$ and 2$b$ at about 1.6 degrees in the observed
CCFs are very probably due to an X--ray source just unresolved that happens to
fall between two of our strips. If the simulations explained above are repeated
with 0.2~deg bins, that peak is also seen in 5 out of a 100 simulations. A
similar effect (due to the overlapping of the long sides of the collimator in
every two adjacent strips) is seen in the poisson CCFs.

{}From the results in Table 2, the UGC sample yields an X--ray volume
emissivity
$j_0=(0.74\pm0.08\pm0.13)\times 10^{39}\thinspace
h\ {\rm erg\ s^{-1}\ Mpc^{-3}}$ while the IRAS one gives
$j_0=(1.15\pm0.10\pm0.19)\times
10^{39}\thinspace h\ {\rm erg\ s^{-1}\ Mpc^{-3}}$. These values are in good
agreement with the zero--lag results obtained by Lahav \etal (1993). The errors
given are the 1 and 2$\sigma$ confidence levels obtained with bootstrap
simulations: for each angular bin in $W_{\rm xg}$ we have
extracted at random (and replaced) as many pairs $(\delta I, \delta N)$ as
in the real data and calculated $\langle\delta N\cdot\delta I\rangle_\theta
/ \langle N\rangle\langle I\rangle$. This was repeated 1000 times and the
resulting CCF was fitted to Eq. 9 to obtain $j_0$; the quoted errors enclose
68.3 per cent (1 sigma) and 95.4 per cent (2 sigma) of the simulations. They
also span the range of values of $j_0$ in Table 2 obtained for each sample,
reflecting thus not only the statistical errors in our fits, but also
uncertainties in our model parameters ($R_{\rm min}$, $\gamma$, $r_0$).

If we had ignored the clustering term (as did Jahoda \etal\ 1991),
we would have obtained $j_0\sim 8\times 10^{39}\thinspace
h\ {\rm erg\ s^{-1}\ Mpc^{-3}}$ for UGC, showing the importance of taking
it into account.

The value found by Miyaji \etal\ (1994) ($j_0=(0.9\pm0.2)\times
10^{39}\thinspace h\ {\rm erg\ s^{-1}\ Mpc^{-3}}$) is in good agreement
with our IRAS result (they overlap within 1 sigma), taking into account the
different IRAS sample (our 0.7~Jy versus their 2~Jy with a local cutoff), the
different instrument (\Ginga~LAC versus \HEAO1~A2), the slightly different
observation bands (4--12~keV versus 2--10~keV) and the different techniques
(full angular and collimator resolution versus zero--lag square collimator)
used.

\beginsection 6 IMPLICATIONS FOR THE X--RAY BACKGROUND

\noindent In this section we study the fraction of the XRB contributed by the
sources whose emissivity we have just obtained from their CCF with optical and
IR galaxies.

The intensity received from objects with local volume emissivity $j_0$
distributed up to redshift $z_{\rm max}$ is

$$I(<z_{\rm max})={\Omega_{eff}\over 4\pi}{c\over H_0} j_0
f(z_{\rm max},\Omega_0,p,\alpha) \eqno(11)$$
where $f$ is the effective look--back factor (Boldt 1982, Lahav 1992) which
depends on the cosmology ($\Omega_0$=1), on the X--ray spectrum of the sources
(we assume a power law with energy index $\alpha=0.7$ which is consistent  with
the XRB fluctuations studied by Butcher \etal 1994), on the evolution of the
emissivity (comoving $j(z)=j_0(1+z)^p$) and, of course, on the maximum
redshift of integration $z_{\rm max}$. Under this assumptions, we get
$$I(<z_{\rm max})={\Omega_{eff}\over 4\pi}{c\over H_0} j_0
\int_{0}^{z_{\rm max}}dz\,(1+\Omega_0 z)^{-1/2}(1+z)^{-2-\alpha+p}\eqno(12)$$

The fraction of the XRB contributed by those objects is then
$F\equiv I(<z_{\rm max})/\langle I\rangle$. In Figs. 3a and 3b we show
contour plots for
different values of
$F$ (for $j_0=0.74$ and $1.15\times10^{39}\thinspace h\ {\rm erg\ s^{-1}
\ Mpc^{-3}}$, UGC and IRAS best fits respectively) in the $(z_{\rm max},p)$
space.
We can see that between 10--20 per cent of the XRB
could be produced if we extrapolate the obtained emissivity to $z_{\rm max}=1$,
and 20--30 per cent if $z_{\rm max}=4$, in the absence of evolution ($p=0$).
Some
moderate positive evolution would lead to much higher values (but see below).
Note that Eq. 12 only depends on $-\alpha+p$, not on their individual
values, so if a different spectral slope $\alpha'$ is considered, the
corresponding $F$ would be given at $(z_{\rm max},p'=p-\alpha'+0.7)$.

The population of sources whose emissivity has been obtained would also produce
some auto--correlation signal on their X--ray intensities (ACF). Any estimation
of their contribution to the XRB must be consistent with the current upper
limits on  the ACF.  We have derived (Fig. 4) the non poissonian ACF of the
X--ray sample in the same way as Carrera \etal (1993) (i.e., by subtracting
from the  observed ACF that expected from a purely poissonian distribution of
sources, obtained by simulations). Again, only upper limits to the cosmic
signal are obtained, a null value being the best estimate of the non
poissonian ACF.

If we assume an evolution for the spatial correlation function of
$\xi(r,z)=(1+z)^{-3-\epsilon}(r/r_0)^{-\gamma}$, then under the small angle
approximation, the non poissonian ACF is given by

$$ \eqalign{W_{xx}({\bf n}_1-{\bf n}_2)={c\over H_0}
&\biggl( {j_0\over 4\pi\langle I\rangle} \biggr) ^2 r_0^\gamma H_\gamma
\int d^2\Omega_n\ \eta_{\rm col}({\bf n}-{\bf n}_1)
\mid {\bf n}-{\bf n}_2\mid^{1-\gamma}\cr
&\int_{0}^{z_{\rm max}}dz\,(1+\Omega_0 z)^{-1/2}(1+z)^{-5-\epsilon-2\alpha+2p}
d_A^{1-\gamma}(z).\cr} \eqno(13)$$
Or, taking into account Eq. 11
$$\eqalign{W_{xx}({\bf n}_1-{\bf n}_2)={H_0\over c}
&{F^2\over \Omega_{eff}^2 f^2(z_{\rm max},\Omega_0,p,\alpha)}
r_0^\gamma H_\gamma
\int d^2\Omega_n\ \eta_{\rm col}({\bf n}-{\bf n}_1)
\mid {\bf n}-{\bf n}_2\mid^{1-\gamma}\cr
&\int_{0}^{z_{\rm max}}dz\,(1+\Omega_0 z)^{-1/2}(1+z)^{-5-\epsilon-2\alpha+2p}
d_A^{1-\gamma}(z),\cr} \eqno(14)$$
where $d_A(z)$ is the angular distance. In this expression, for a given
pair $(z_{\rm max},p)$, the only free parameter is $F$, the fraction of the
XRB produced by the objects whose ACF we are studying.

To get a quantitative idea of the constraints that this ACF places on the
possible values of $p$ and $z_{\rm max}$, we have $\chi^2$--fitted the observed
non--poissonian ACF (Fig. 4) to Eq. 14
over the range $0.2-5$ deg, with $F$ as a the only free parameter. The
best fit (under the current assumptions on the clustering properties of the
sources) always corresponds to $F=0$; for each pair $(z_{\rm max},p)$ the
maximum
allowed value of $F$ (from $\Delta\chi^2=4$ or 2$\sigma$)
is plotted in Figs. $5a$ and $5b$ for comoving and
stable clustering ($\epsilon=-1.2$ and $\epsilon=0$, respectively) and
for UGC ($\gamma=1.8$ and $r_0=5 \hmpc$). Figs. $5c$ and
$5d$ correspond (respectively) to the same evolutions of the clustering
for IRAS ($\gamma=1.8$ and $r_0=4 \hmpc$). Combining Figs. 3 and 5
we can see that the ACF upper limits do not allow contributions to the
XRB greater than $\sim 40-50$~per cent, even with evolution.

In Section 4 we assumed that $j_{\rm xg}=j_0$. If we keep both quantities
separate, $j_0$ (the total emissivity) is
replaced by $j_{\rm xg}$ (the emissivity of the sources in the catalogue)
as the free quantity to fit, and
$r_0$ by $r{_0}'=r_0(j_0/j_{\rm xg})^{1/\gamma}>r_0$. Hence, the net effect of
only a fraction of the galaxies being X--ray emitters would be to enhance
the importance of the clustering term with respect to the poisson term (as
expected).
As the clustering term is already dominant for the UGC sample, reducing
$j_{\rm xg}/j_0$ even down to 25~per cent does not have any effect on the
derived
total $j_0$. However, the poisson term is dominant for the IRAS sample and as
we
decrease $j_{\rm xg}/j_0$ the clustering term takes over. This, in turn,
increases
the total emissivity (by $\sim$50~per cent when $j_{\rm xg}/j_0=0.25$), but the
ACF constraints keep $F\approxlt$20~per cent.

The difference between the X--ray emissivities obtained for the UGC and IRAS
samples is likely to reflect an intrinsic difference in properties of their
constituent galaxies. For example, the UGC catalogue selects galaxies
by their diameter, without distinction of type, while the IRAS catalogue is
a far--infrared selected sample, with a high proportion of star forming
and dust rich galaxies, hence mostly spirals.

We know that sources producing the {\sl Ginga} fluctuations have a spectrum
with energy index $\sim 0.7$ (Butcher \etal 1994), and these are precisely the
sources whose emissivity has been obtained here. With only a moderate
contribution to the XRB, the residual spectrum of the XRB would be much flatter
than the presently observed one (energy index $\sim0.4$). This will have  wide
implications for the sources of this residual XRB, which will need to be highly
absorbed and/or reflection dominated.

\beginsection 7 CONCLUSIONS

\noindent X--ray observations of the 2--10~keV XRB in scan mode performed with
the {\sl
Ginga} satellite have been cross--correlated with several catalogues of
different objects (mainly galaxies). The significance level of the resulting
CCF signals have been established using simulations of both X--ray and
catalogue
samples.

We have concluded that there is a strong cross--correlation between the hard
XRB and nearby galaxies (especially those in UGC and the IRAS catalogue or
Meurs \& Harmon 1989), finding $W_{\rm xg}(0)\sim1.2-1.4\times 10^{-2}$.

Expressions have been derived for the expected CCF at non--zero lag when the
population of X--ray sources dominating the {\it Ginga} fluctuations are
associated or strongly clustered with the catalogued galaxies. In the specific
case of the X--ray sources being present in the galaxy catalogue,
least--squares fitting has been used to obtain the X--ray volume emissivity
$j_0$ necessary to produce the observed signal, obtaining $j_0\sim0.7-1.2\times
10^{39}\thinspace h\ {\rm erg\ s^{-1} \ Mpc^{-3}}$.

Extrapolating this emissivity back to $z=1$ in the absence of evolution, 10--20
per cent of the XRB would be produced ($\approxlt$30 per cent if the
contribution from the non--evolving sources is integrated up to $z=4$). We have
shown that these fractions are consistent with the absence of any ACF signal in
the XRB.

Higher fractions, close to saturation, would be achieved even with very
moderate evolution. However, these high fractions are not permitted by the
upper limits on the ACF. Furthermore, the spectrum of
the summed contribution of these sources would be very different from the
observed XRB spectrum, unless rapid spectral evolution also takes place.

Direct observations with higher angular resolution instruments
({\sl ASCA}, SPECTRUM--X--$\Gamma$, {\sl XMM}) would be very useful to decide
on whether the emissivities obtained here are actually from the sources in
the catalogues, or come from X--ray sources clustered with them. They would
also provide spectra for fainter sources, helping to decipher the riddle
of the hard X--ray background.

\beginsection ACKNOWLEDGEMENTS

\noindent XB thanks the DGICYT for financial support, under project PB92-00501.
FJC, XB and ACF acknowledge partial financial support from a NATO collaborative
research grant. XB and ACF were partly supported by the ``Human Capital and
Mobility'' programme of the EU under contract CHRX-CT92-0033.

\beginsection REFERENCES

\ref Abell, G.O., 1958. ApJS, 3, 211

\ref Barcons, X., Branduardi--Raymont, G., Warwick, R.S., Fabian, A.C., Mason,
K.O., Mc Hardy, I., Rowan--Robinson, M., 1994. MNRAS, 268, 833

\ref Boldt, E.,1987. Phys. Reports, 146, 215

\ref Branduardi--Raymont, G., Mason, K.O., Warwick, R.S., Carrera, F.J.,
Mittaz, J.P.D., Puchnarewicz, E.M., Smith, P.J., Barber, C.R., Pounds, K.A.,
Stewart, G.C., M$^{\rm c}$Hardy, I.M., Jones, L.R., Merrifield, M.R., Fabian,
A.C., M$^{\rm c}$Mahon, R., Ward, M.J., George, I.M., Jones, M.H., Lawrence,
A., Rowan--Robinson, M., 1994. MNRAS, 270, 947

\ref Butcher J.A., \etal, 1994. Submitted to MNRAS

\ref Carrera, F.J., Barcons, X., Butcher, J.A., Fabian, A.C., Stewart, G.C.,
Toffolatti, L., Warwick, R.S., Hayashida, K., Inoue, H., Kondo, H., 1993.
MNRAS, 260, 376

\ref Danese, L., De Zotti, G., Andreani, P., 1992. In:  The X--Ray
Background, ed. Barcons, X., Fabian, A.C., Cambridge Univ. Press, p. 61

\ref Fabian, A.C., Barcons, X., 1992. ARA\&A,  30, 429

\ref Hasinger, G., Burg, R., Giacconi, R., Hartner, G., Schmidt,, M.,
Tr\" umper, J., Zamorani, G., 1993. A\&A, 271, 1

\ref Grossan, A.G., 1992. PhD Thesis, Massachusetts Institute of
Technology

\ref Huchra, J.P., Davis, M., Latham, D., Tonry, J., 1983. ApJS, 52,89

\ref Hudson, M.J., Lynden--Bell, D., 1991. MNRAS, 252, 219

\ref Jahoda, K., Lahav, O., Mushotzky, R.F., Boldt, E.A.,  1991. ApJ,
378, L37.
Erratum in 1992. ApJ, 399, L107

\ref Kondo, H., 1990. PhD Thesis, University of Tokyo

\ref Lahav, O., 1992. In:  The X--Ray Background, ed. Barcons, X. \&
Fabian, A.C., Cambridge Univ. Press, p. 102

\ref Lahav, O., Fabian, A.C., Barcons, X., Boldt, E., Butcher, J., Carrera,
F.J., Jahoda, K., Miyaji, T., Stewart, G.C., Warwick, R.S., 1993.
Nat, 364, 693

\ref Marshall, F.E., Boldt, E.A., Holt, S.S., Miller, B.B., Mushotzsky, R.F.,
Rose, L.A., Rothschild, R.E., Serlemitsos, P.J.,  1980. ApJ, 235, 4

\ref Meurs, E.J.A., Harmon, R.T., 1989. A\&A,  206, 53

\ref Miyaji, T., Lahav, O., Jahoda, K., Boldt, E., 1994. ApJ, 434, 424

\ref Nilson, P.,  1973. {\it Uppsala General Catalogue of Galaxies}, Uppsala
Astr. Obs. ann., 6

\ref Piccinotti, G. Mushotzky, R.F., Boldt, E.A., Holt, S.S., Marshall, F.E.,
Serlemitsos, P.J., Shafer, R.A.,  1982. ApJ, 253, 485

\ref Rowan--Robinson, M., Lawrence, A., Saunders, W., Crawford, J., Ellis, R.,
French, C.S., Parry, I., Xiaoyang, X., Allington--Smith, J., Efstathiou, G.,
 Kaiser, N., 1990. MNRAS, 247, 1

\ref Shanks, T., Georgantopoulos, I., Stewart, G.C., Pounds, K.A., Boyle, B.J.,
 Griffiths, R.E.,  1991. Nat, 353, 315

\ref Strauss, M.A., Davis, M., Yahil, A., Huchra, J.P., 1990. ApJ, 361, 49

\ref Yahil, A., Strauss, M.A., Davis, M., Huchra, J.P., 1991.
ApJ,  372, 380

\vfill\eject

\beginsection FIGURE CAPTIONS

\noindent {\bf Figure 1.} $W_{\rm xg}(\theta)$
in 1~degree bins (filled
dots): the error bars are extracted from the standard deviation in each bin.
Also shown are the median of the simulations (solid line) and the limits
encompassing 68 (dashed lines) and 95 (dotted lines) per cent of the
simulations. $a$ is for the UGC sample and $b$ for the IRAS sample (see text).

\noindent {\bf Figure 2.} $W_{\rm xg}(\theta)$ in
0.2~degree bins (filled dots): the error bars are extracted from the
simulations. Also shown are the best total fit (solid line) and the
clustering (dashed line) and poissonian terms (dotted line). $a$ is for the
UGC sample and $b$ for the IRAS sample. The peaks at
about 1.6 degrees present in both observed CCFs
are due to a just unresolved X--ray source between observed sky strips (see
text).

\noindent {\bf Figure 3.} Contour plots (in the ($z_{\rm max}$,$p$) space)
of the fraction of the X--ray
background $F$ produced by a population with local emissivity $j_0$: levels
of F=0.1 (solid line), 0.3 (dashed line),  0.5 (dot--dashed line), 0.7 (dotted
line) and 0.9 (dash--dot--dot--dot line) are shown.
$a$ is for $j_0=0.74$ (UGC) and $b$ for $j_0=1.15$ (IRAS), both in units of
$10^{39}\thinspace h\ {\rm erg\ s^{-1}\ Mpc^{-3}}$.

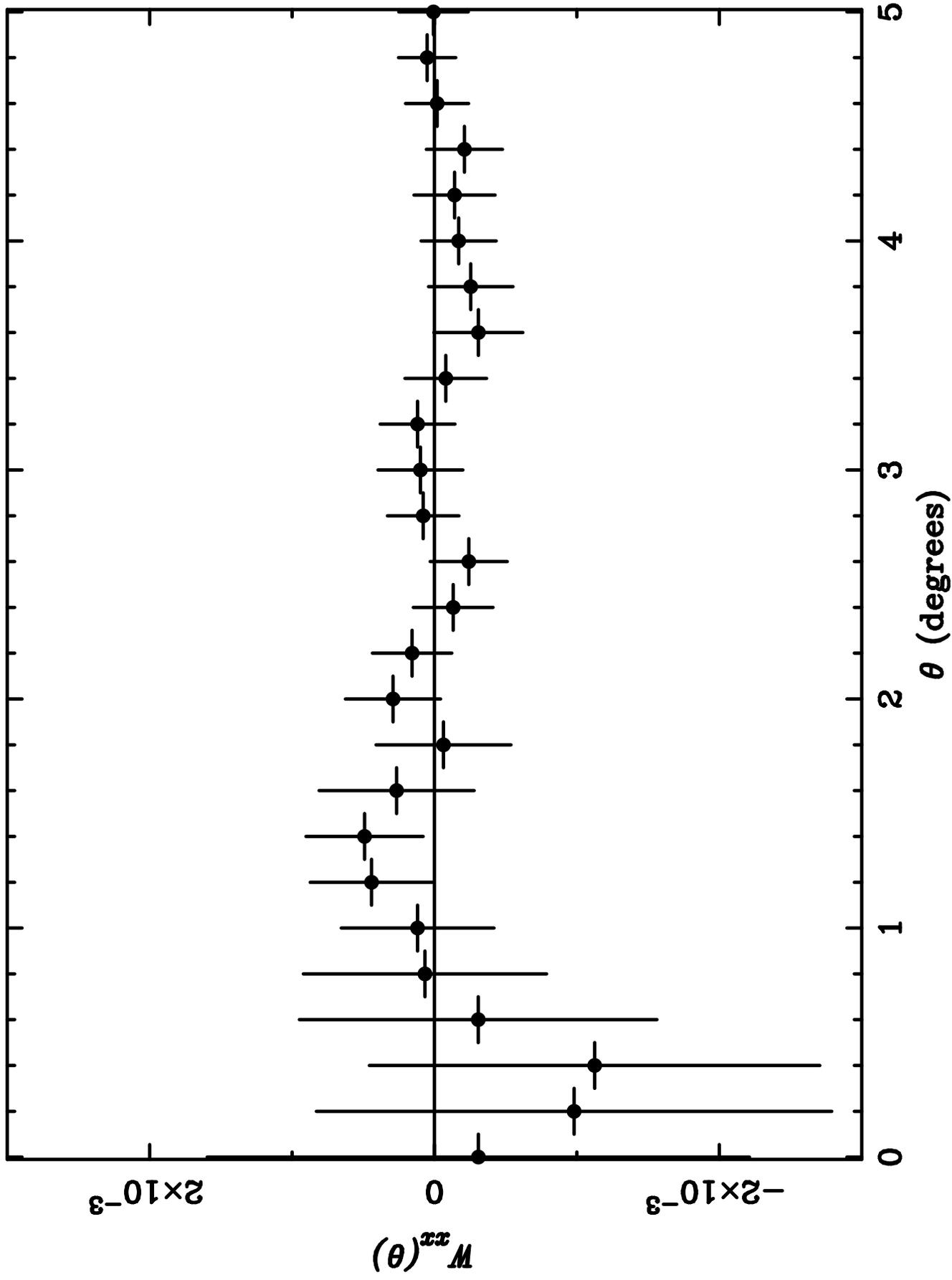
\noindent {\bf Figure 4.} Non--poissonian ACF of the XRB: the error bars
($1\sigma$) are extracted from simulations (see text).

\noindent {\bf Figure 5.} Contour plots of the maximum $F$ allowed by the
ACF. Levels
of F=0.1 (solid line), 0.3 (dashed line),  0.5 (dot--dashed line), 0.7 (dotted
line) and 0.9 (dash--dot--dot--dot line) are shown.  $a$ is for comoving
clustering and UGC, $b$ for stable clustering and UGC, $c$ is for comoving
clustering and IRAS and $d$ for stable clustering and IRAS.

\vfill\eject

\noindent{\bf Table 1.} Cross--correlation functions or several samples with
the XRB (see text).

$$\vbox{\halign{
\hfil#\hfil & \hfil#\hfil & \hfil#\hfil &  \hfil#\hfil       & \hfil#\hfil \cr
\noalign{\hrule}
\noalign{\smallskip}
 Name       &\ N\         &\ \  $\langle N\rangle$\ \ &\ \ $W_{\rm xg}(0)$\ &
             \  $W_{\rm xg,sim}(0)$\cr
\noalign{\smallskip}
\noalign{\hrule}
\noalign{\smallskip}
UGC         & 951         & 1.24        & 0.0116$\pm$0.0015  & $<$0.008 \cr
IRAS        & 351         & 0.48        & 0.0142$\pm$0.0016  & $<$0.010 \cr
IRASQ0      & 534         & 0.74        & 0.0063$\pm$0.0013  & $<$0.009 \cr
CfA         & 196         & 0.32        & 0.0190$\pm$0.0030  & $<$0.018 \cr
Abell       & 210         & 0.26        & 0.0100$\pm$0.0020  & $<$0.017 \cr
IRASZ       &  82         & 0.10        & 0.0240$\pm$0.0050  & $<$0.026 \cr
IRASZS      &  45         & 0.06        & 0.0460$\pm$0.0080  & $<$0.026 \cr
IRASZSVOL   &  45         & 1.97        & 0.0060$\pm$0.0060  & $<$0.069 \cr
\noalign{\smallskip}
\noalign{\hrule}
}}$$

\noindent {\bf Table 2.} Results of the fitting for different parameters
(see text).

$$\vbox{\halign{
#\hfil & \hfil#\hfil &\hfil#\hfil &   \hfil#\hfil  & \hfil#\hfil  &
              #\hfill \cr
\noalign{\hrule}
\noalign{\smallskip}
Sample & $r_0$       & $\gamma$   &\ Range fitted\ &\ $R_{\rm min}\ $ &
              \ \ \ \ \ $j_0\,(\pm1\sigma\pm2\sigma)$              \cr
       &\ ($\hmpc$)\ &            & (degrees)      &\ ($\hmpc$)\  &
              ($10^{39}\thinspace h {\rm\,erg\,s^{-1}\,Mpc^{-3}}$)\ \cr
\noalign{\smallskip}
\noalign{\hrule}
\noalign{\smallskip}
UGC       & 5.00 & 1.80 & 0.2,2.0 & 11.1 & 0.82                     \cr
UGC       & 5.00 & 1.80 & 0.2,5.0 & 11.1 & 0.65                     \cr
UGC       & 5.00 & 1.80 & 0.2,2.0 & 13.3 & 0.92                     \cr
UGC*      & 5.00 & 1.80 & 0.2,5.0 & 13.3 & 0.74 ($\pm 0.07\pm0.13$) \cr
\noalign{\smallskip}
\noalign{\hrule}
\noalign{\smallskip}
IRAS      & 4.00 & 1.80 & 0.2,2.0 &\ 1.0 & 1.22                    \cr
IRAS      & 4.00 & 1.80 & 0.2,2.0 &\ 5.0 & 1.23                    \cr
IRAS      & 4.00 & 1.80 & 0.2,2.0 & 10.0 & 1.24                    \cr
IRAS      & 4.00 & 1.80 & 0.2,5.0 &\ 1.0 & 1.14                    \cr
IRAS*     & 4.00 & 1.80 & 0.2,5.0 &\ 5.0 & 1.15 ($\pm0.10\pm0.20$) \cr
IRAS      & 4.00 & 1.80 & 0.2,5.0 & 10.0 & 1.17                    \cr
IRAS      & 4.00 & 1.65 & 0.2,2.0 &\ 5.0 & 1.34                    \cr
IRAS      & 4.00 & 1.65 & 0.2,5.0 &\ 5.0 & 1.23                    \cr
IRAS      & 3.75 & 1.65 & 0.2,2.0 &\ 5.0 & 1.40                    \cr
IRAS      & 3.75 & 1.65 & 0.2,5.0 &\ 5.0 & 1.30                    \cr
\noalign{\smallskip}
\noalign{\hrule}
}}$$

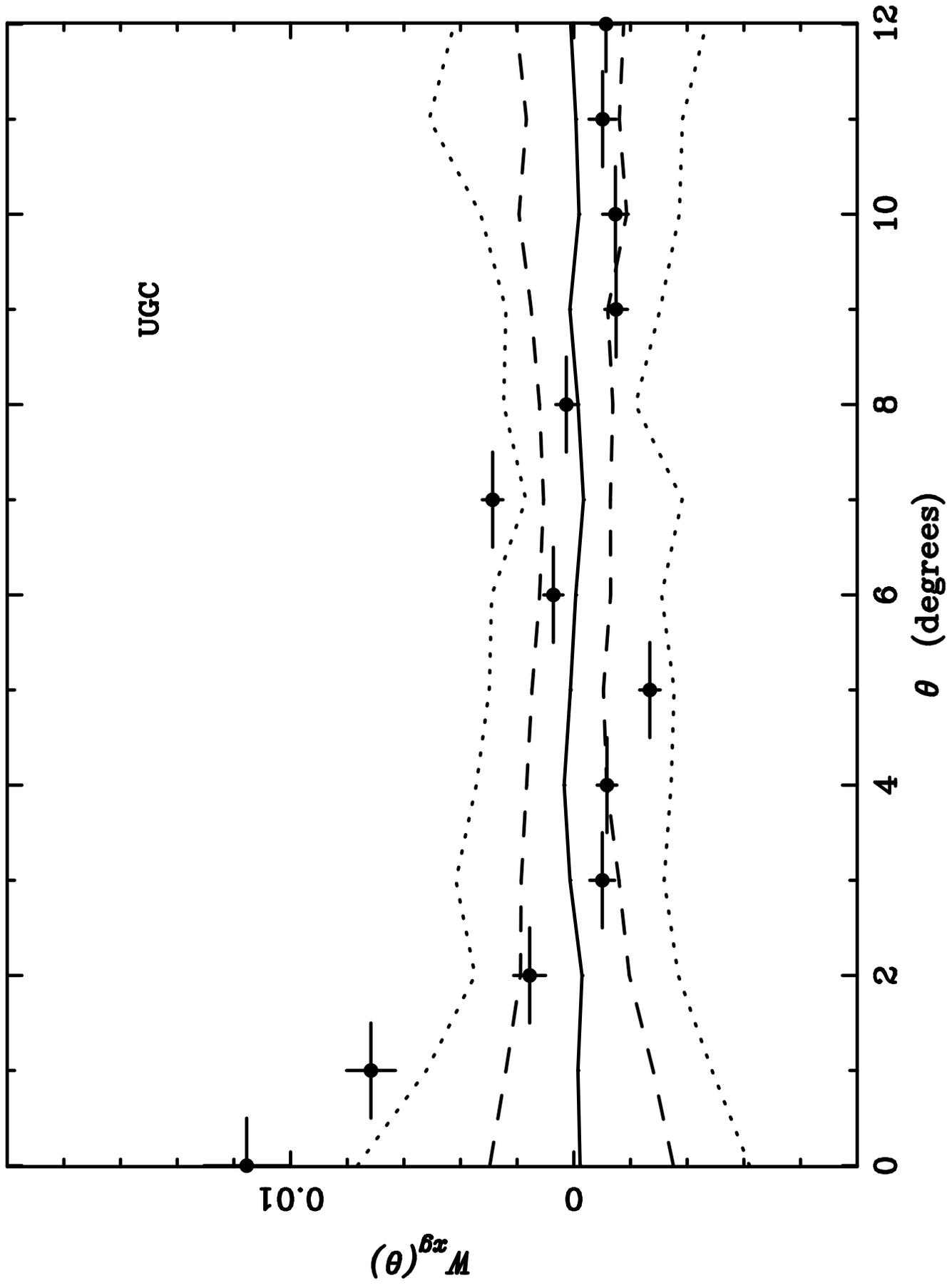
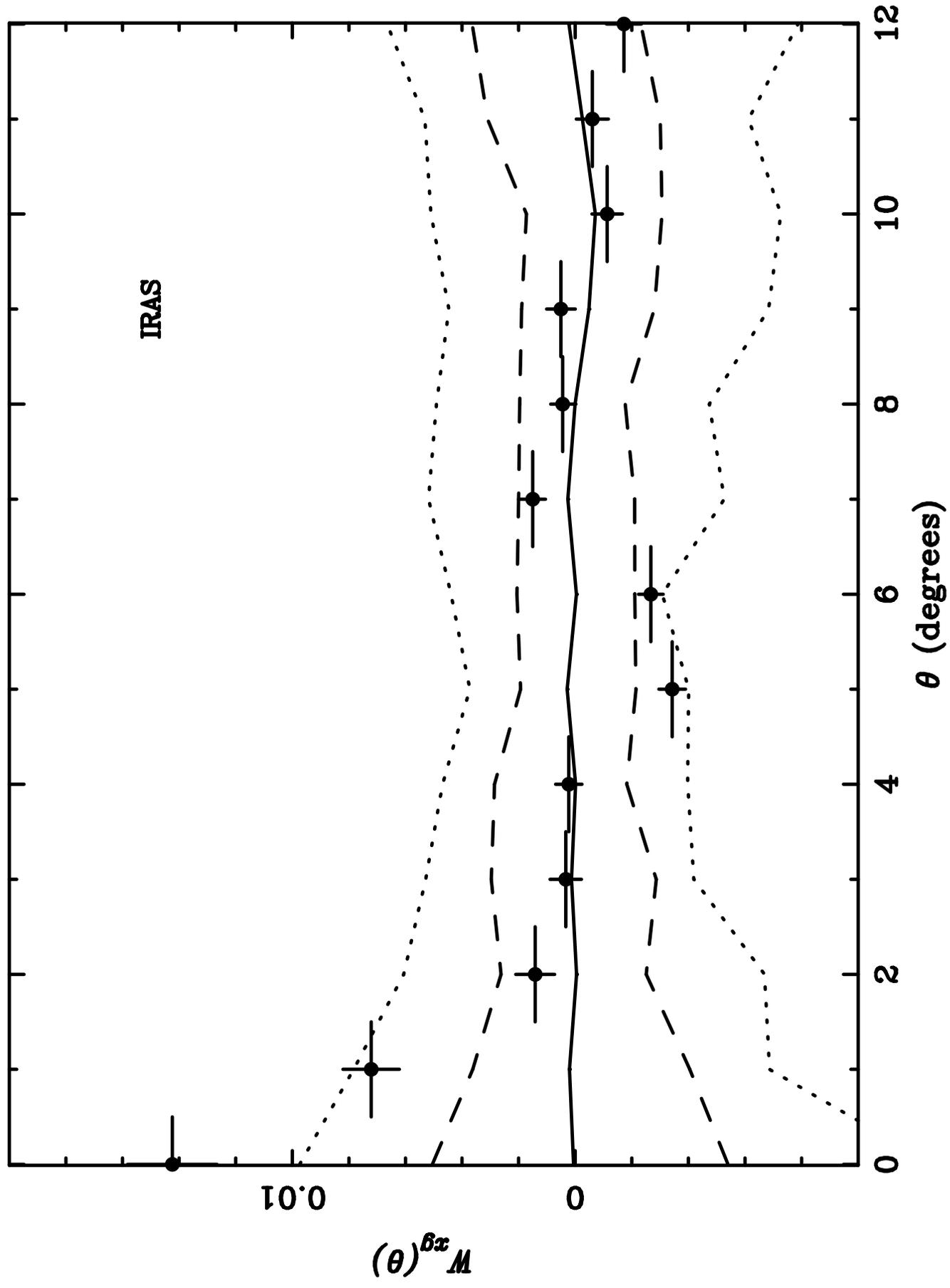
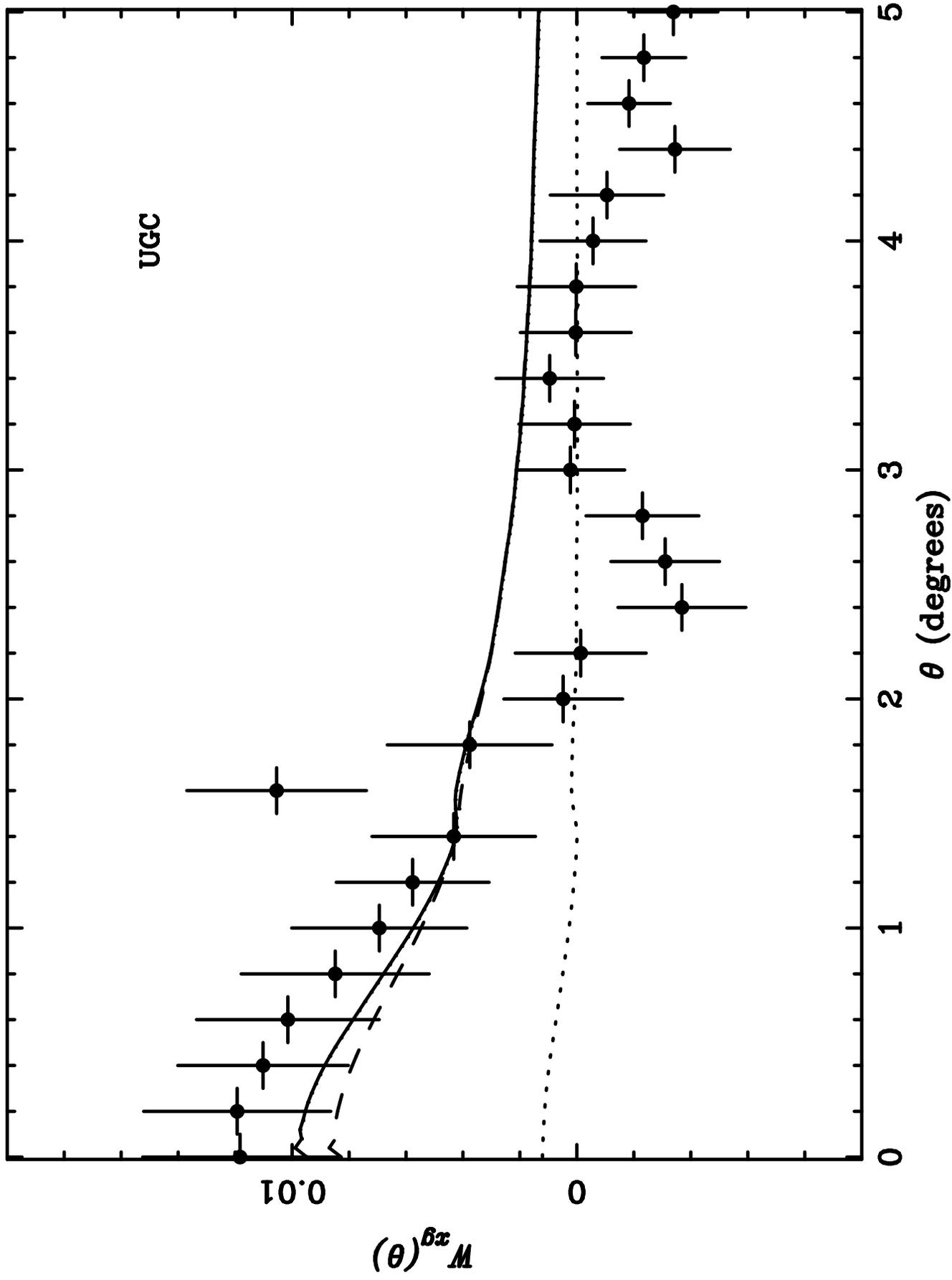
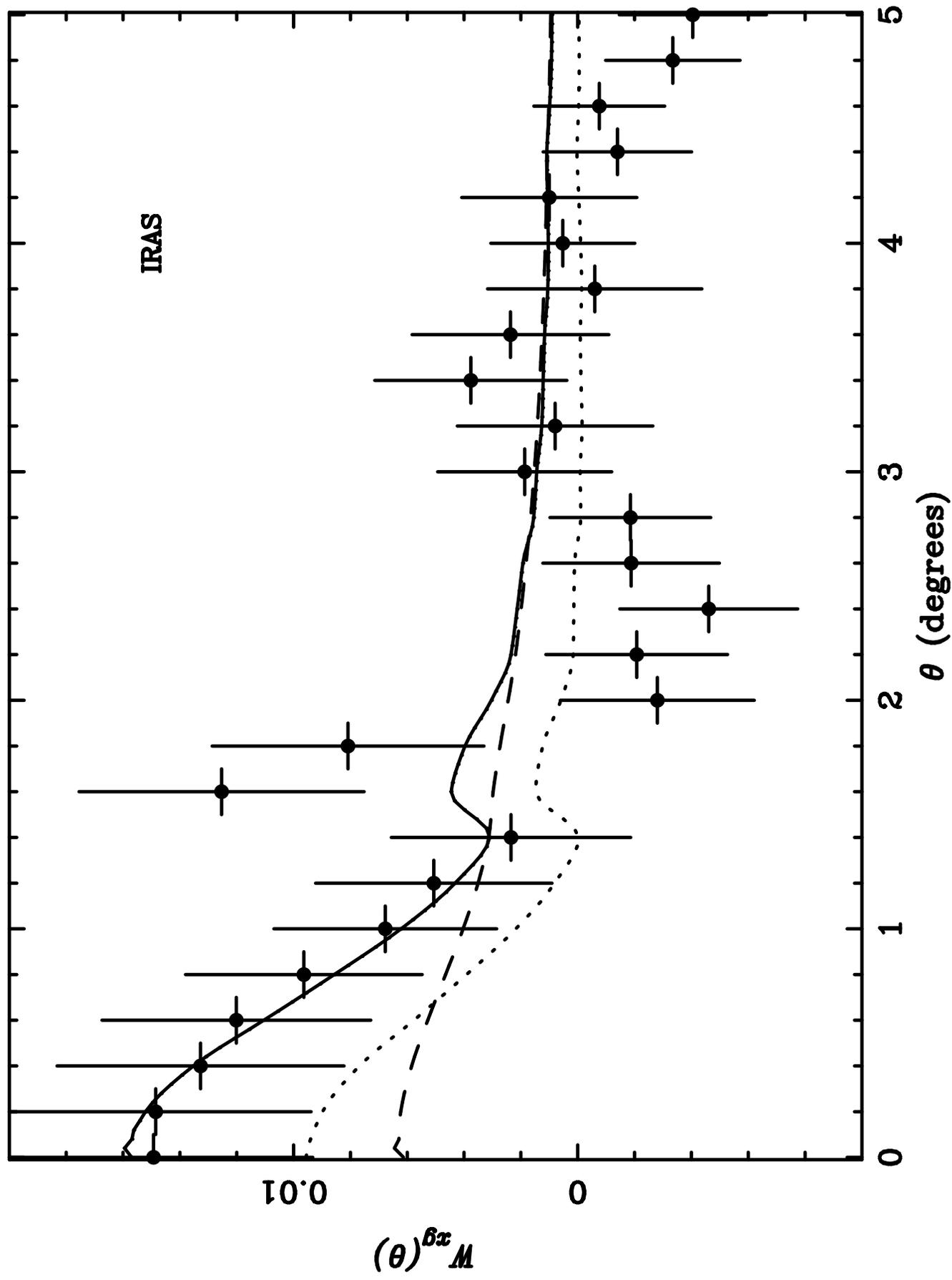
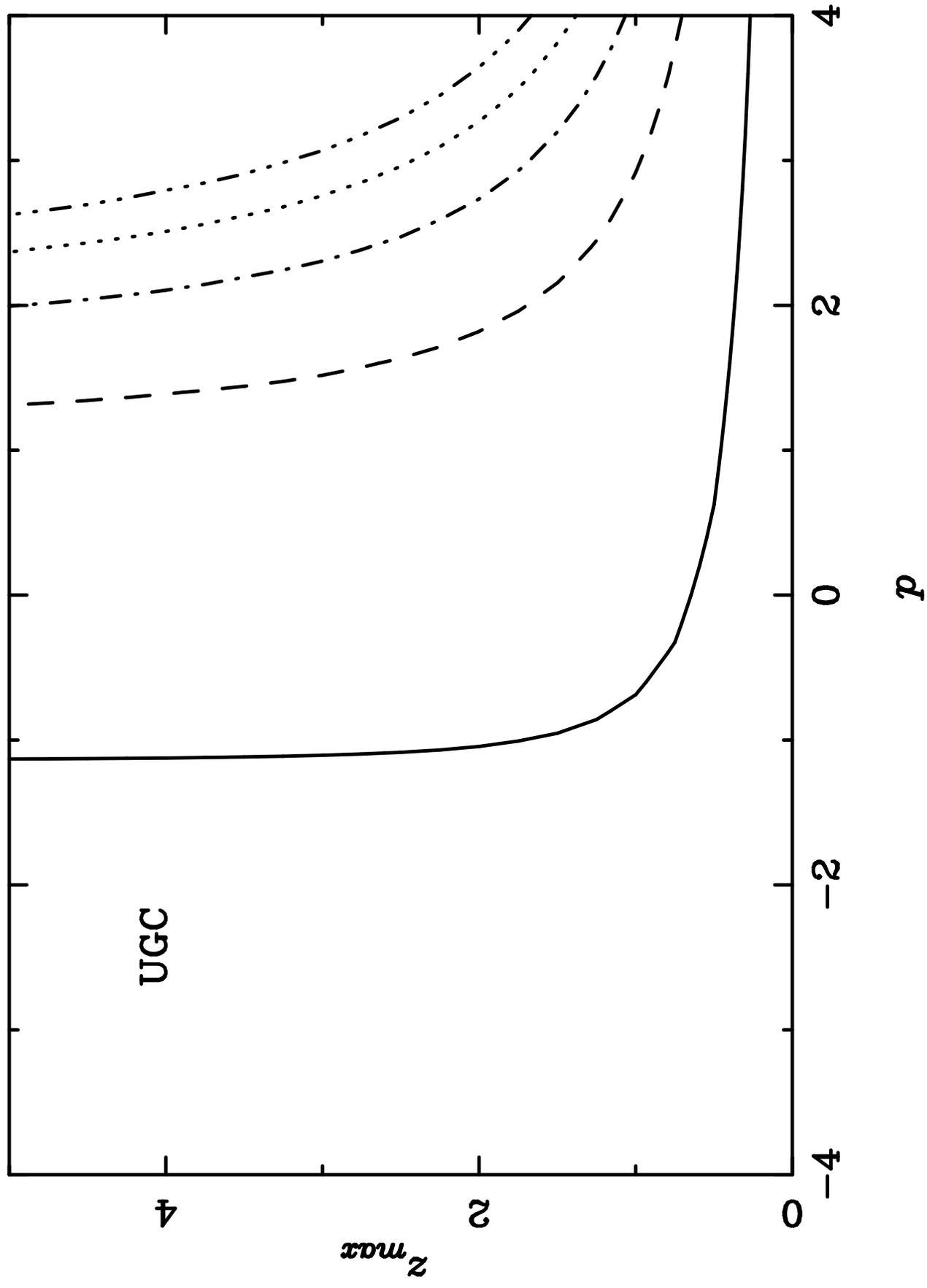
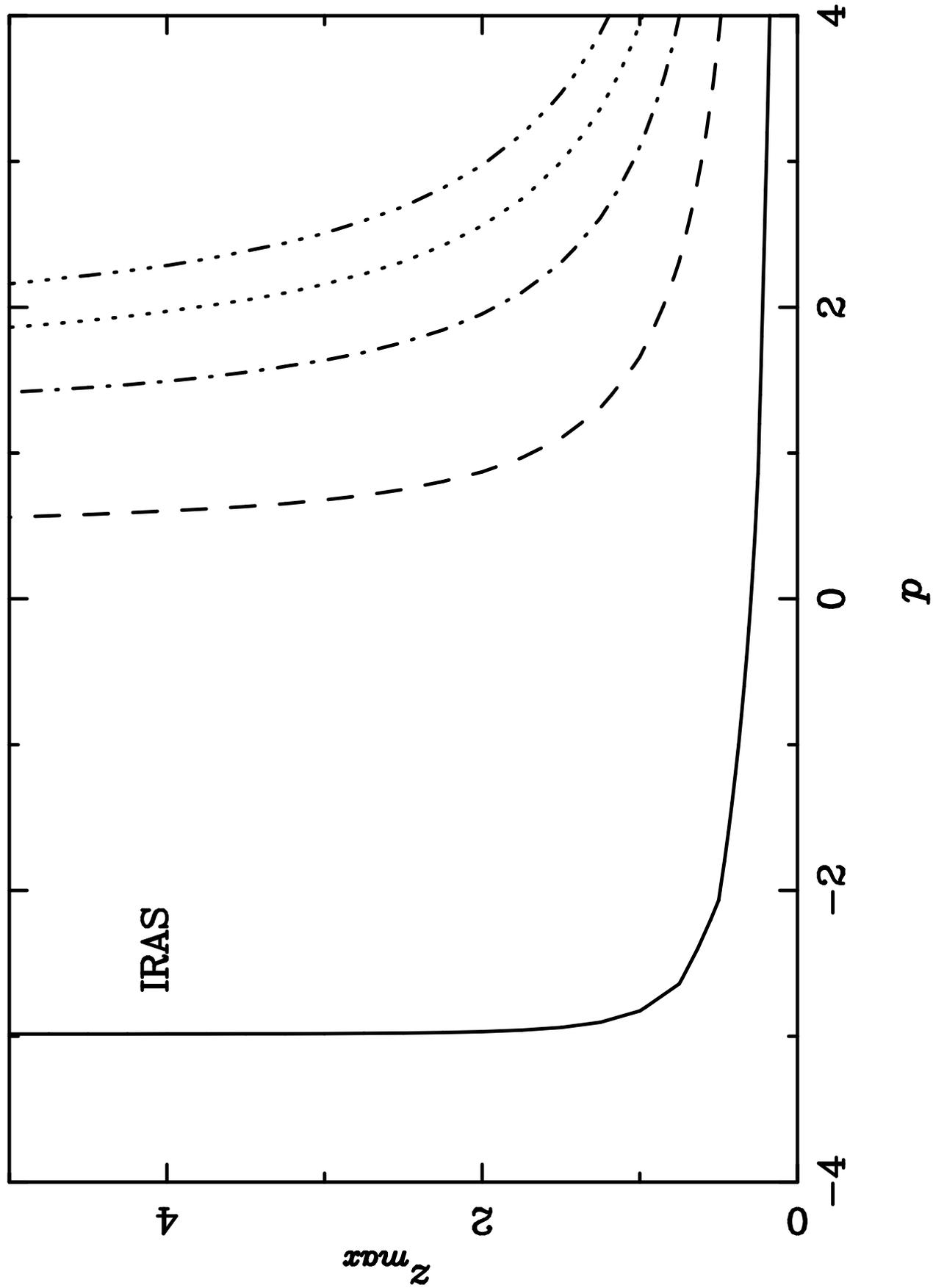
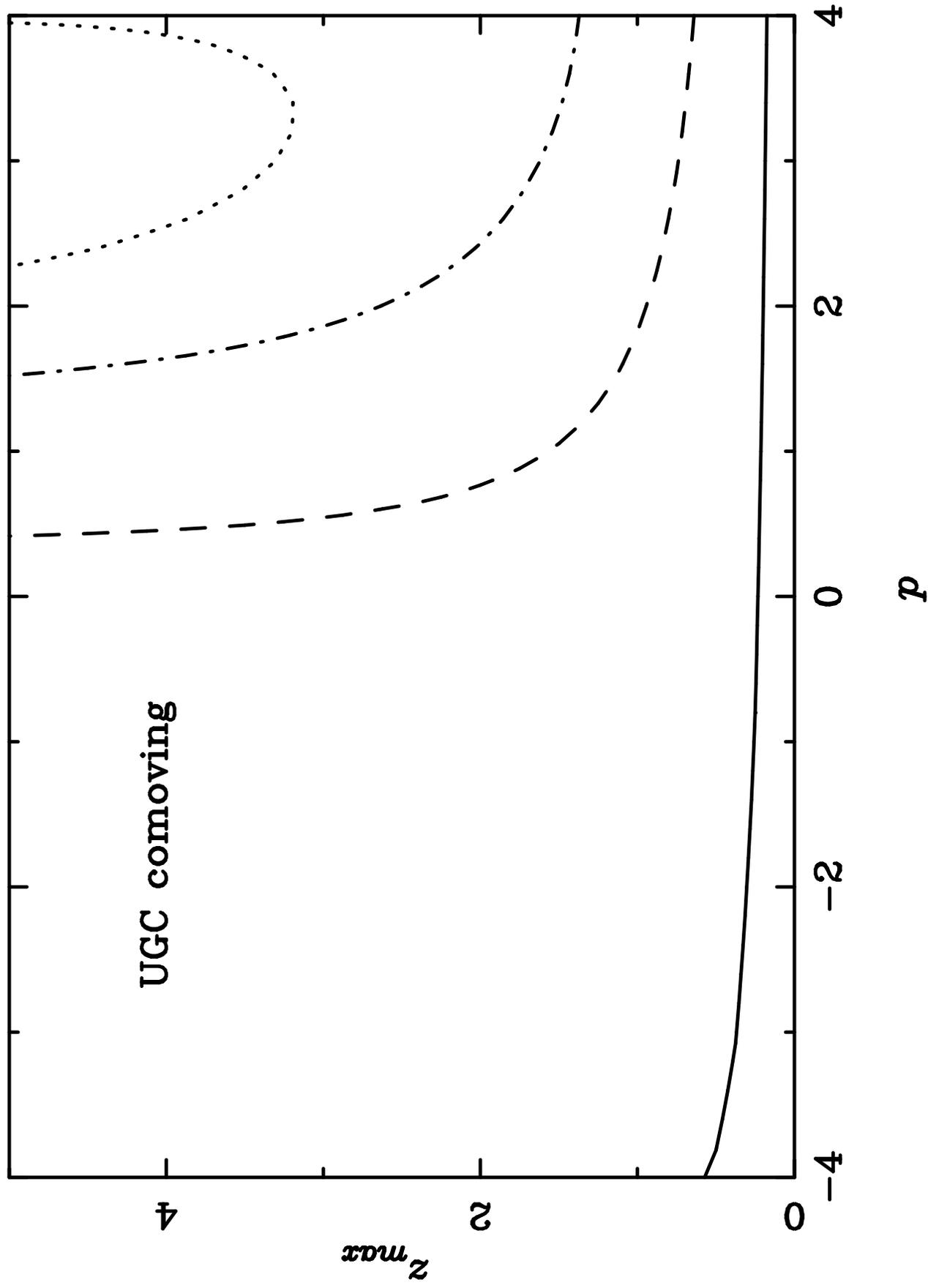
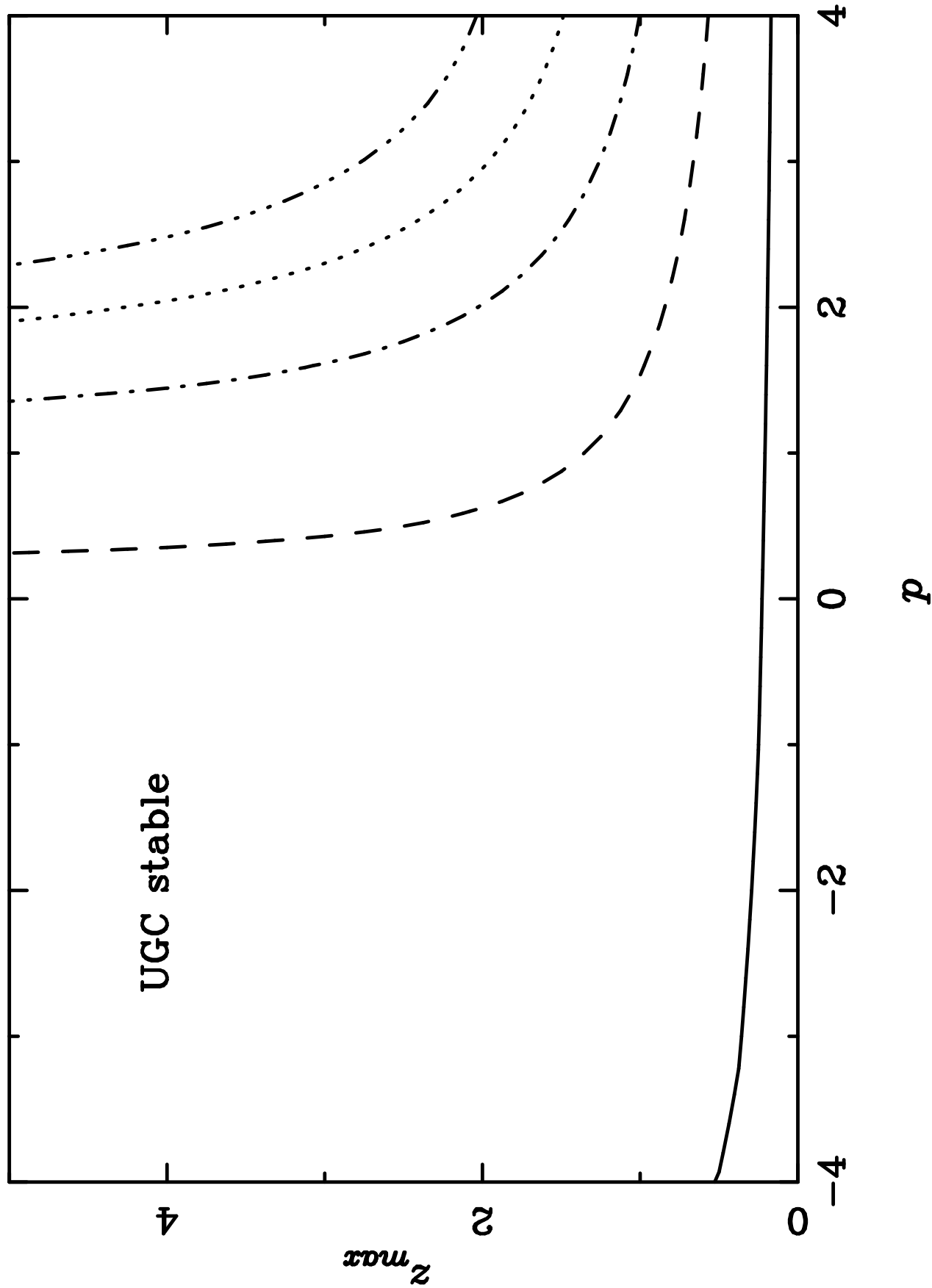
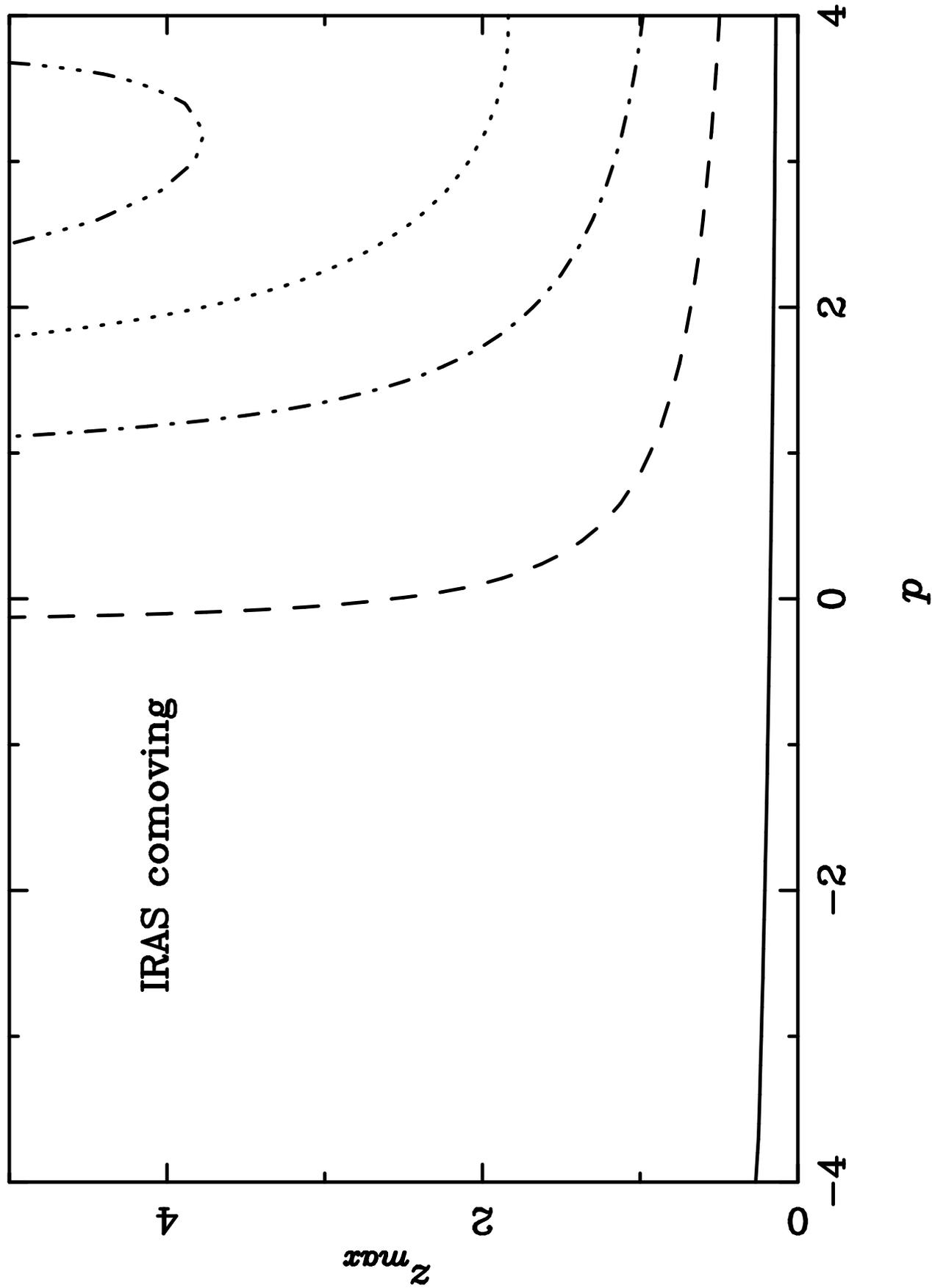
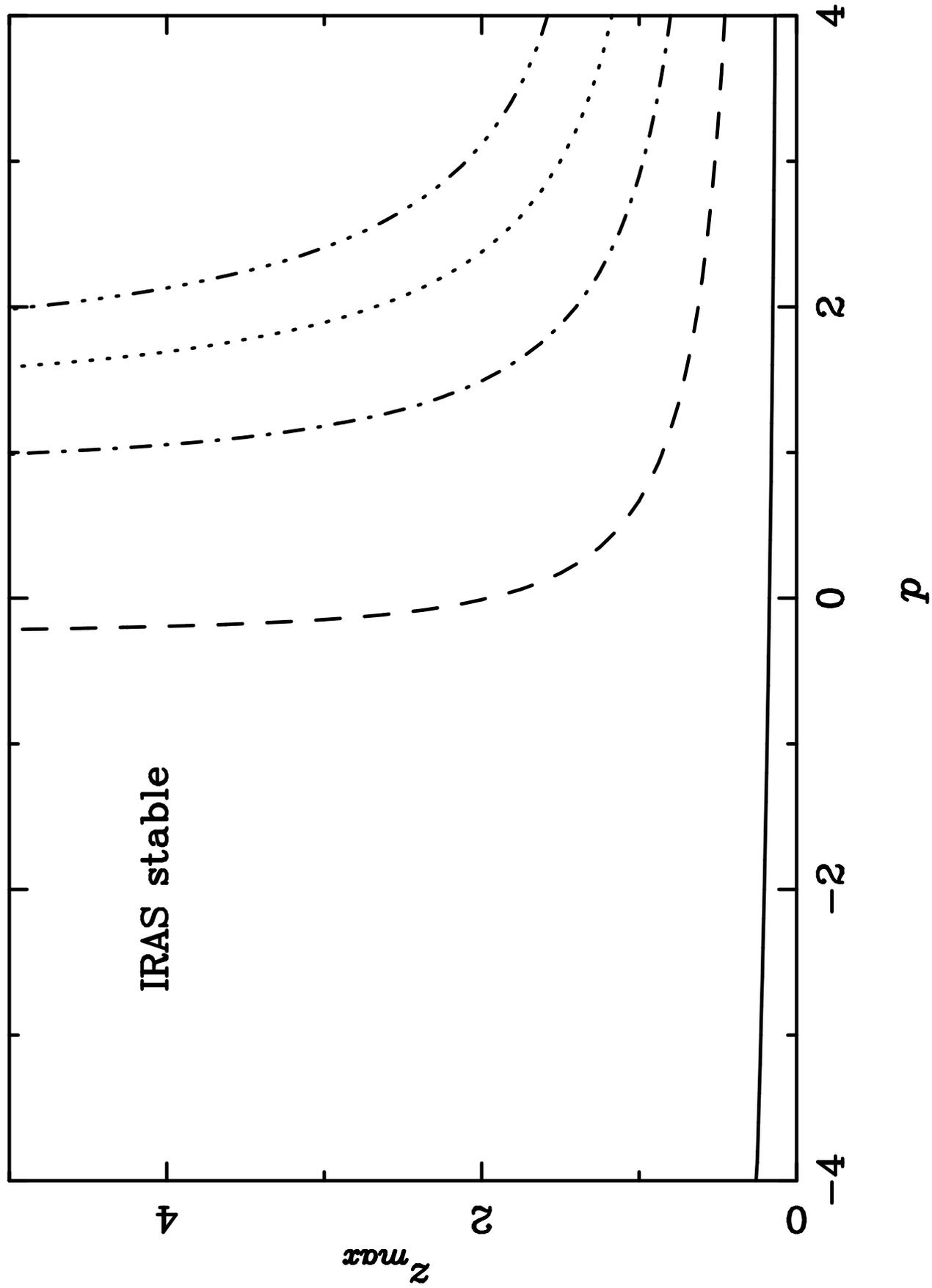
\bye